\documentclass[aps, prb, preprint]{revtex4}
\usepackage{graphicx}

\begin{document}

\title{A first principles study on organic molecules encapsulated BN nanotubes }

\author{Wei He}
\author{Zhenyu Li}
\author{Jinlong Yang}
\thanks{Corresponding author. E-mail: jlyang@ustc.edu.cn}
\author{J. G. Hou}

\affiliation{Hefei National Laboratory for Physical Sciences at
Microscale,  University of Science and Technology of China, Hefei,
Anhui 230026, China}

\date{\today}

\begin{abstract}
The electronic structures of boron nitride nanotubes (BNNTs) doped
by organic molecules are investigated with density functional
theory. Electrophilic molecule introduces acceptor states in the
wide gap of BNNT close to the valence band edge, which makes the
doped system a $p$-type semiconductor. However, with typical
nucleophilic organic molecules encapsulation, only deep occupied
molecular states but no shallow donor states are observed. There is
a significant electron transfer from BNNT to electrophilic molecule,
while the charge transfer between nucleophilic molecule and BNNT is
neglectable. When both electrophilic and nucleophilic molecules are
encapsulated in the same BNNT, large charge transfer between the two
kinds of molecules occurs. The resulted small energy gap can
strongly modify the transport and optical properties of the system.
\end{abstract}


\maketitle

\section{Introduction}

Since its discovery in 1991, \cite{discovery} carbon nanotube (CNT)
has attracted a considerable attention due to its novel properties
and great potential for nanodevice applications. As a structural
analogue to CNT, boron nitride nanotube (BNNT) has distinctive
properties. Due to its large ionicity, BNNT is a wide-gap
semiconductor. \cite{bn1, bn2} More importantly, the energy gap
($\sim$5.5 eV) does not depend on the diameter, chirality, and the
number of walls of the tube. \cite{bn3} On the other hand, BNNT has
excellent mechanical stiffness and thermal conductivity. BNNT also
shows pronounced resistance to oxidation, and it is stable up to 700
$^{\circ}$C in air, while CNT survives only below 400 $^{\circ}$C.
\cite{bn5} All these properties make BNNT an attractive candidate
for nano-electronics.

For electronics applications, it is very desirable to modify the
band structure of BNNT to obtain metallic or $p$-type/$n$-type
semiconducting behavior. Many attempts at this direction have been
reported. The energy gap can be adjusted by changing the chemical
composition, for example, by substituting the B or N atoms with C
atoms. \cite{bcn1, bcn2, bcn3, bcn4} However, it is difficult to
control precisely the atom concentration of carbon within BCN
nanotube at the stage of tube growth.  BNNT band structure can also
be tuned by organic molecules covalent functionalization, \cite{ex0,
ex2, ex, ex1} or by applying transverse electric field through the
Stark effect. \cite{field1, field2}

Another kind of methods for BNNT electronic structure engineering is
encapsulation. Previously investigated compounds encapsulated in
BNNT include metal atoms \cite{metal2} and fullerenes. \cite{C601,
C602, C603} Compared to them, there are many advantages to use
organic molecule encapsulation. Organic molecules are typically air
stable, simple to synthesis, and abundant. They have already been
used to adjust CNT band structure. \cite{organic0} It was found that
the carrier type and concentration in CNT can be easily controled by
tuning the electron affinity (EA) and ionization potential (IP) of
the encapsulated molecules. \cite{organic0, organic1} It is thus
very interesting to study the possible electronic structure
engineering by organic molecule encapsulation in BNNT.

In this paper, we investigate the electronic structures of organic
molecule encapsulated BNNTs (denoted as Molecule@BN) using density
functional theory (DFT). Two kinds of organic molecules are studied:
electrophilic and nucleophilic molecules. The electrophilic
molecules are easy to get electrons, and typically have a large EA.
The two examples of electrophilic molecules studied in this paper
are Tetracyano-p-quinodimethane (TCNQ) and
tetrafluorocyano-p-quinodimethane (F4TCNQ). Their adiabatic EA are
2.80 and 3.38 eV, respectively. \cite{organic0} The nucleophilic
molecules are easy to lose electrons, and typically have small IP.
Three typical nucleophilic molecules are selected:
tetrakis(dimethylamino)ethylene (TDAE), anthracene (ANTR), and
tetrathia-fulvalene (TTF). Their adiabatic IP are 5.36, 6.40, and
7.36 eV, respectively. \cite{organic0} The structure of these
molecules are shown in the top row of Fig. \ref{fig:total}.

\section{Computational Details}

Spin-polarized DFT  were used to investigate the geometric and
electronic structures of the Molecule@BN systems. The calculations
were performed with the Vienna Ab Initio Simulation Package (VASP).
\cite{17,18} We adopted the Perdew, Burke, and Ernzerhof
exchange-correlation functional within the generalized gradient
approximation (PBE-GGA), \cite{19} and the projector augmented-wave
(PAW) pesudopotential. \cite{20,21} 1s orbital of B, N, C, and F and
1s to 2p orbitals of S were treated as core states. The energy
cutoffs used in our calculations for the plane-waves were 400 eV.

The (16,0) BNNT was chosen as an example, which has a diameter of
12.8 \AA, similar to the CNT used in the previous experiment.
\cite{organic0} The periodic boundary condition was empolyed, and
each BNNT was separated by 10 \AA\ of vacuum.  A unit cell
containing 64 B and 64 N atoms was used, with the minimum distance
between periodically repeated organic molecule centers along the
BNNT tube axis to be $\sim$8.7 \AA. This is consistent with the
molecular concentration in previous experiment for CNT.
\cite{organic0} The 4$\times$1$\times$1 $\Gamma$-centered
Monhkorst-Pack $k$-point grid was used for the sampling of the
Brillouin zone.

Geometry optimization was performed by a few steps. First, the
pristine BNNT was optimized. The resulted B-N bond length was 1.451
\AA, agreeing well with the experimental value (1.45 \AA). The
separately relaxed organic molecule was then inserted into the BNNT.
Initially, the organic molecule was put with its center on the tube
axis and with its long axis (C=C double bond direction except ANTR)
making a 27 degree angle with respect to the tube axis. This angle
was chosen to maintain the separate distance between each molecule
large enough. Finally, the molecules in the BNNT were allowed to
fully relax until the forces were less than 20 meV/\AA, and the BNNT
was fixed to its previously optimized geometry. Our test calculation
indicates that simultaneously relaxing the BNNT and molecule does
not affect our results. The optimized geometry is shown in Fig.
\ref{fig:total}. The molecule geometries changed little after
inserted into BNNT, which suggests the weak interaction between the
BNNT and molecules due to the large distance between them.

\section{Results and Discussion}

We define the binding energy of the molecule and the BNNT as
\begin{equation}
E_b=E_{BN}+E_{M}-E_{BN-M}
\end{equation}
where $E_{BN}$ and $E_M$ are the energies of individual BNNT and
molecule, respectively. $E_{BN-M}$ is the energy of the Molecule@BN
system. The calculated binding energies for the five different
molecules are listed in Table \ref{tbl:eb}. As we can see, for all
molecules, $E_b$ is positive, which means that the molecules can be
inserted into BNNT from the thermodynamical viewpoint. On the other
hand, the absolute values of the binding energy are relatively
small. This is consistent with the weak interaction between the BNNT
and the encapsulated molecules.

\begin{table}[b]
\caption{Binding energies and charge transfers between molecule and
BN for the Molecule@BN systems. Energies are in meV per molecule.
Charge transfers are in electron per molecule, negative sign
corresponds to electron from molecule to BNNT. } \label{tbl:eb}
\begin{tabular}{cccccc}
 \hline\hline
Molecule & TCNQ & F4TCNQ & TDEA & TTF & ANTR  \\
 \hline
$E_b$  & 92.6 & 162.6 & 71.4  & 65.8  & 73.5   \\
$q_t$  & 0.10  & 0.24  & -0.004  & -0.008  & -0.005   \\
 \hline\hline
\end{tabular}
\end{table}

The band structures of the pristine and doped BNNT are plotted in
Fig. \ref{fig:band0}. The calculated band gap for the pristine BNNT
is 4.5 eV, which is a little smaller than the experiment value due
to the tendency of DFT to underestimate band gap. The band structure
of the doped BNNT can be roughly considered as a simple combination
of the band structures of the BNNT and the inside organic molecules,
due to the weak interaction between them. There are small
dispersions for some molecular bands, which is due to
molecule-molecule interaction as suggested by our charge density
analysis (see Fig. \ref{fig:STM}a). For electrophilic organic
molecules, their lowest unoccupied molecular orbital (LUMO) forms
the lowest unoccupied band of the Molecule@BN system, and the
highest occupied band comes from BNNT. The new energy gap is 0.16
and 0.10 eV for TCNQ and F4TCNQ, respectively. Therefore, doping of
electrophilic molecules turns BNNT to $p$-type semiconductor.

For the three nucleophilic molecules, their highest occupied
molecular orbital (HOMO) corresponds to the highest occupied band of
the doped BNNTs. For TDAE@BN, the lowest unoccupied band is the
conduction band edge of the BNNT, and the new energy gap is 2.22 eV.
For TTF@BN and ANTR@BN, the lowest unoccupied band comes from the
LUMO of the doped molecule, and the energy gap is 1.95 and 2.26 eV,
respectively. The energy gaps are too large to make the nucleophilic
organic molecule doped BNNT to be a good $n$-type semiconductor.

This behavior contrasts with that of CNT, where both $n$-type and
$p$-type semiconductors can be realized by organic molecule
encapsulation. \cite{organic0, organic1} This difference is caused
by the energy gap width difference between BNNTs and semiconducting
CNTs. The conduction band edge of BNNT is $\sim$1.93 eV below the
vacuum energy level, which means that a dopant with IP close to 1.93
eV may form $n$-type semiconducting BNNT. Such small IP values are
typically not available in organic molecules. We expect that some
metal atom dopings may convert BNNT to $n$-type semiconductor.

We note that, besides transport property engineering, organic
molecule doping will strongly modify the optical property of BNNT,
as suggested by previous study on CNT. \cite{organic2} The gap of
pristine BNNT is too large to develop possible optical application
for it. With molecule doping, the gap width can be adjusted in a
large extent, which gives the possibility to develop
photo-electronic device based on BNNT.

Another interesting property is the charge transfer between doped
molecules and the BNNT. The charge transfer between covalently
connected functional groups and BNNT has been studied previously.
\cite{ex2} To calculate the charge transfer, we follow the procedure
proposed by Lu et al.,\cite{charge1} in which an cylinder-integrated
differential electron density curve against the radial coordinate is
plotted in order to determine the boundary between the organic
molecule and the nanotube. We suppose the boundary between the
molecule and the BNNT is a cylinder with radius $r_b$. This is a
good approximation, as suggested by Fig. \ref{fig:STM}b. Before we
plot the cylinder-integrated differential electron density curve, we
would like to estimate the reasonable range of $r_b$. For the
electrophilic organic molecules, electrons transfer from the valence
band of BNNT to the LUMO of molecule, so the boundary should be
inside the main density distribution of the highest occupied band of
BNNT ($r_b < 6$ \AA), as shown in Fig. \ref{fig:STM}c. For the
nucleophilic organic molecules, electrons transfer from the HOMO of
molecule to the conduction band of BNNT, so the boundary should be
inside the lowest unoccupied band of BNNT. Since the conduction band
edge of the BNNT is a nearly-free-electron state, which extend to
the far inner part of the tube, the boundary radius should be
relatively small for nucleophilic molecules ($r_b < 3.5$ \AA), as
shown in Fig. \ref{fig:STM}d.

As suggested by Lu et al.,\cite{charge1} the exact value of the
boundary radius $r_b$ is finally defined as the maximum/minimum
position of the cylinder-integrated differential electron density
curve (see Fig. \ref{fig:E0.0charge}), and the value of the charge
transfer is corresponding maximum/minimum value. Differential
electron density is the electron density difference between the
Molecule@BN system and separated molecule and BNNT. For TCNQ@BN and
F4TCNQ@BN, we have a maximum peak at 4.4 \AA. Therefore, the
boundary radial position $r_b$ is 4.4 \AA, which is roughly the
midpoint between the molecule and the BNNT. The calculated charge
transfer for BNNT to TCNQ and F4TCNQ are 0.10 and 0.24 electrons per
molecule, respectively. The charge transfers between the
nucleophilic molecules and BNNT are more complicated. There are
several comparable maximum and minimum peaks in the
cylinder-integrated differential density curve. According to the
estimation of the range of $r_b$, we set the position of the first
minimum as the boundary between molecule and BNNT. The corresponding
boundary radial position $r_b$ are 2.8, 2.8, and 3.0 \AA\ for TDAE,
TTF, and ANTR, respectively. The resulted charge transfer are very
small. We note that for the three nucleophilic molecules, the charge
transfer is always very small no matter where the boundary is set.

Since we have high occupied band for nucleophilic molecule doping,
and low unoccupied band for electrophilic molecule doping, it is
thus interesting to study the possibility to dope both nucleophilic
and electrophilic molecules into the same BNNT. The bulk material
composed by TCNQ and TTF has been widely studied, and it is an
interesting one-dimensional conductor. \cite{Sing0311} There is a
large charge transfer about 0.59 electrons between TCNQ and TTF. The
electronic structure of TCNQ-TTF@BN is studied, where the unit cell
was doubled compared to single-type molecule doping. One TCNQ and
one TTF molecules was put in the unit cell within the BNNT. Geometry
optimization was performed with the BNNT fixed. The resulted binding
energy $E_b$ is 0.68 eV per unit cell. This large $E_b$ mainly
contributed by the interaction between TCNQ and TTF.

The calculated band structure of TCNQ-TTF@BN is plotted in Fig
\ref{fig:band2}. Compared to band structure of TCNQ@BN and TTF@BN,
the TCNQ bands are upshifted, and the TTF bands are downshifted.
This is driven by a 0.43 electrons charge-transfer between TCNQ and
TTF in BNNT. However, the charge transfer between the molecules and
BNNT is neglectable. The lowest unoccupied band form TCNQ and the
highest occupied band from TTF from a very small band gap (67 meV)
for TCNQ-TTF@BN. Such a small gap demonstrates the great potential
of molecular doping to modify the optical property of BNNT. As shown
in Fig \ref{fig:band2}, combing other electrophilic and nucleophilic
molecules gives similar results. The gap becomes larger when IP of
of the nucleophilic molecules becomes larger.

To examine the role of BNNT in the ground-state electronic structure
of TCNQ-TTF@BN, we calculate the electronic structure of the
TCNQ-TTF molecular chain without BNNT, where the geometry of the
molecular chain is fixed to their geometry in BNNT. Such model
system reproduces the bands originated from molecules in
TCNQ-TTF@BN. This result indicates that the electronic structure
close to the Fermi level is not affected by BNNT. With excellent
thermo-stability and mechanic properties, the inner space of BNNT
thus provides a good place to develop charge transfer salt based
one-dimensional molecular wire.

\section{Conclusions}

In summary, we calculate the electronic structure of BNNT with
different molecule doping. We can obtain $p$-type semiconductor by
inserting electrophilic molecules. While for typical nucleophilic
organic molecules doped BNNTs, their gap are too large to be a good
$n$-type semiconductor, and the corresponding charge transfers are
also very small. This is because that the energy gap of BNNT is too
large. We expect that some metal atom dopings may convert BNNT to
$n$-type semiconductor. Based on the large extent of the band gap
modification, we expect that the optical property of BNNT will be
strongly modified by molecule doping. When electrophilic and
nucleophilic molecules are doped into BNNT at the same time, a very
small gap formed by a pair of molecular bands is obtained. The
position of these two band within the BNNT gap and their spacing can
be controlled by IP and EA of the molecular dopant, which leads to a
powerful technique to engineer the electronic and optical properties
of BNNTs.

\begin{acknowledgements} This work is partially supported by the
National Natural Science Foundation of China (50121202, 20533030,
20628304), by National Key Basic Research Program under Grant
No.2006CB922004, by the USTC-HP HPC project, and by the SCCAS and
Shanghai Supercomputer Center.
\end{acknowledgements}

\clearpage

\begin{figure}
\caption{Sketch drawing of the organic molecules and the geometries
of the Molecule@BN systems.} \label{fig:total}
\end{figure}

\begin{figure}
\caption{Band structures (from $\Gamma$ to X) of the pristine BNNT
and Molecule@BN. Dashed lines indicate the Fermi levels.}
\label{fig:band0}
\end{figure}

\begin{figure}
\caption{(a) Isosurface of the density of the lowest unoccupied
molecular band of TCNQ@BN. The density iso-value is 0.001 e/\AA$^3$.
(b) Isosurface of the charge density of TTF@BN. The iso-value of
electron density is 0.11 e/\AA$^3$. (c) the highest occupied band
and (d) the lowest unoccupied band of BNNT. } \label{fig:STM}
\end{figure}

\begin{figure}
\caption{Curves of cylinder-integrated differential electron density
against the radial coordinate in doped BNNTs. The boundary between
molecules and BNNT is indicated by a blue line.}
  \label{fig:E0.0charge}
\end{figure}

\begin{figure}
\caption{Geometry (top) and band structures (bottom, from $\Gamma$
to X) of the Molecule@BN system, where electrophilic and
nucleophilic molecules are encapsulated in the same BNNT. The
position of Fermi level is shifted to zero.} \label{fig:band2}
\end{figure}

\clearpage
\begin{center}
\includegraphics[width=12cm]{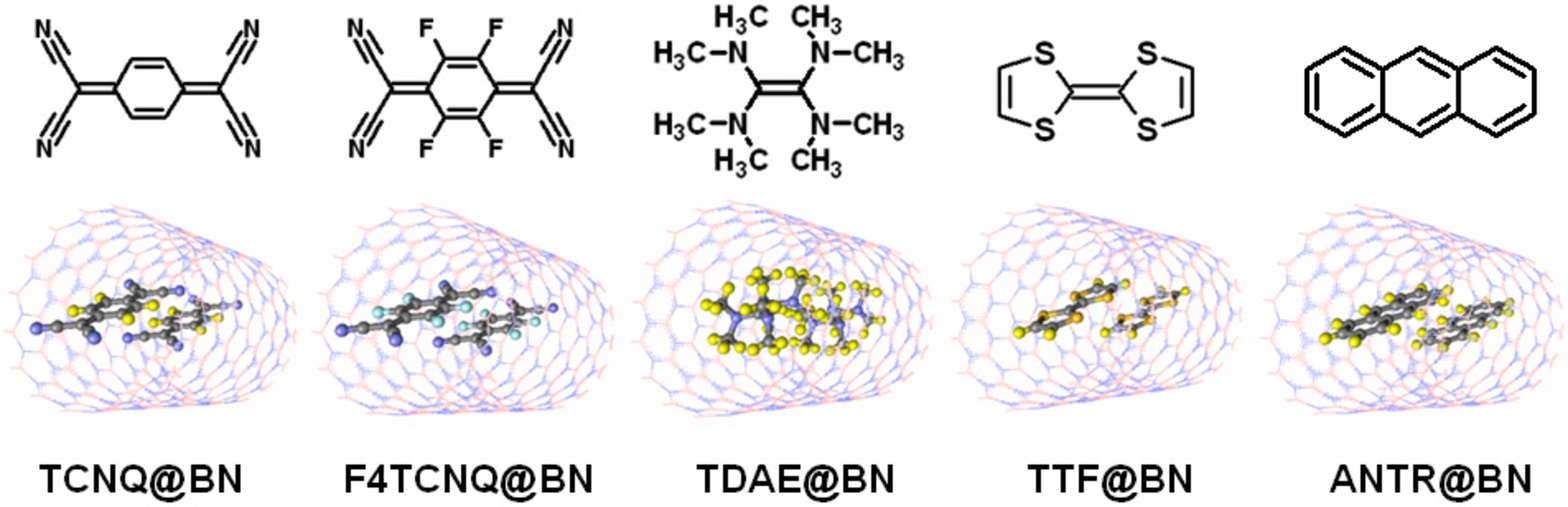} \\
\vspace{5cm} {\large He et al., JCP, Figure 1}
\end{center}

\clearpage
\begin{center}
\includegraphics[width=12cm]{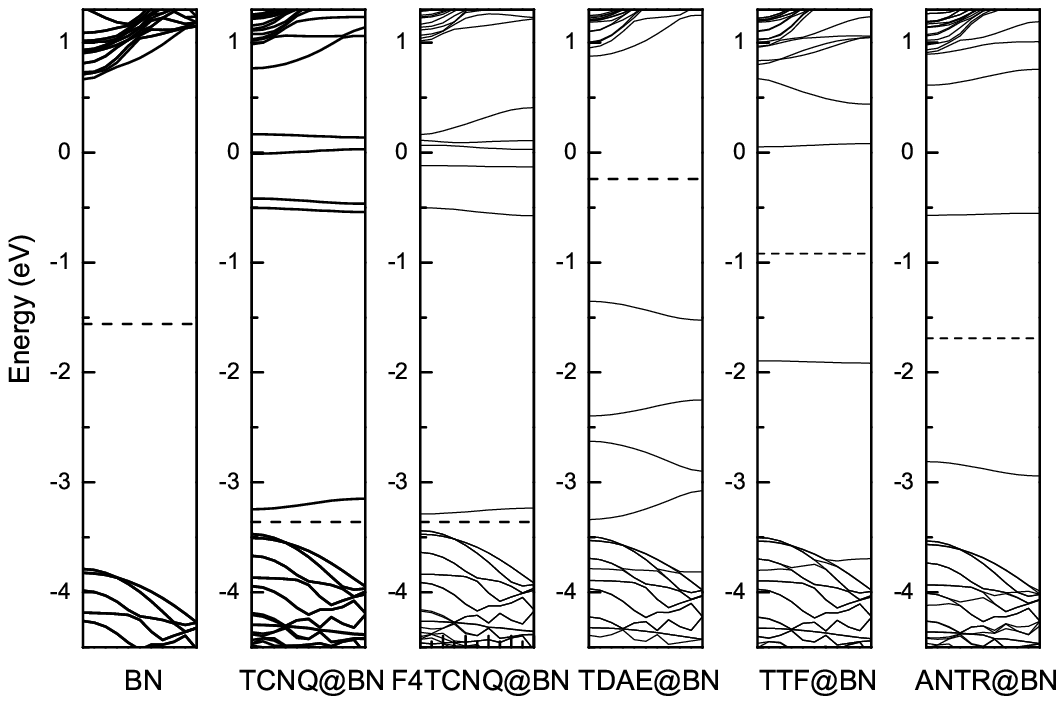} \\
\vspace{5cm} {\large He et al., JCP, Figure 2}
\end{center}

\clearpage
\begin{center}
\includegraphics[width=9cm]{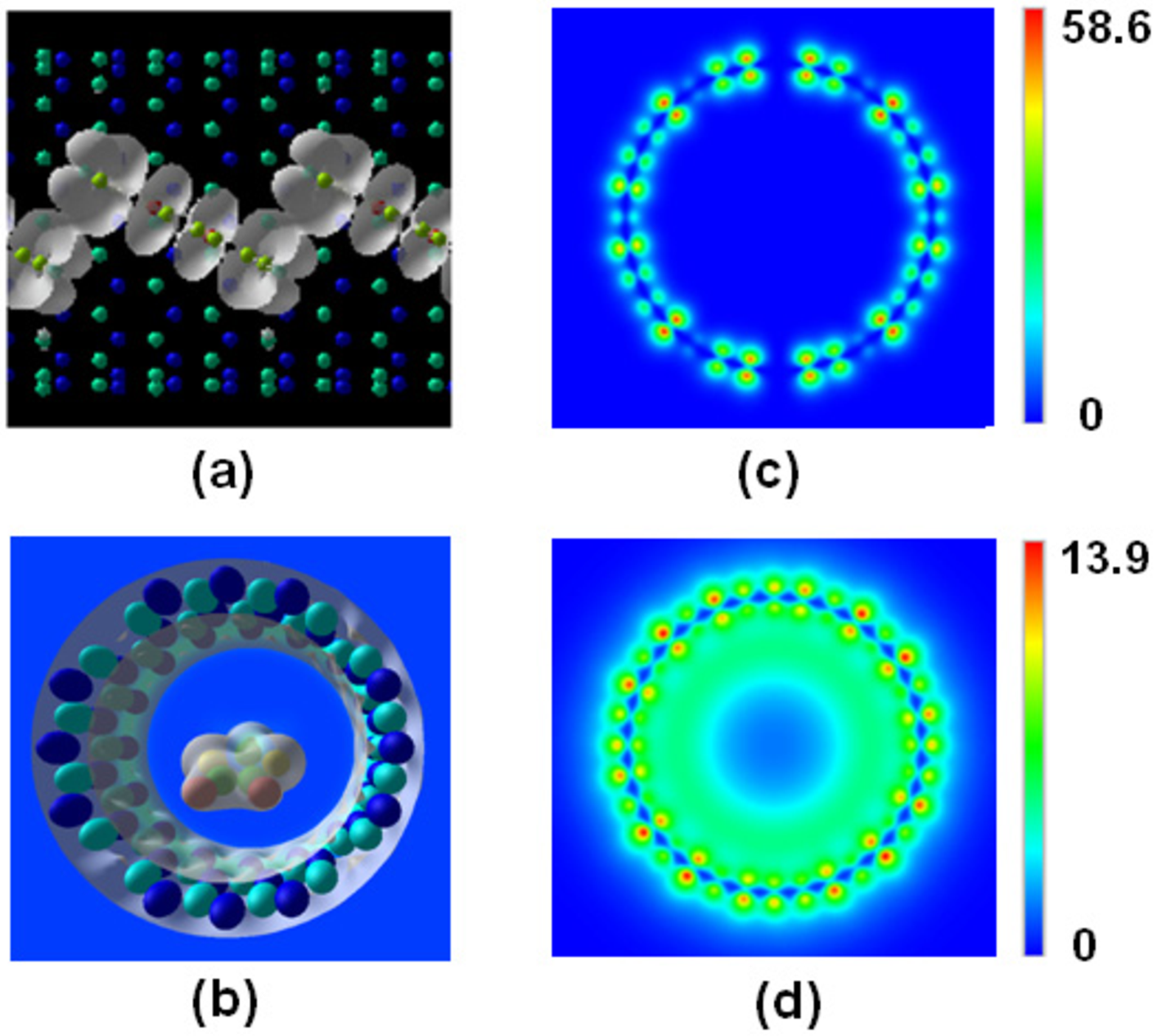} \\
\vspace{5cm} {\large He et al., JCP, Figure 3}
\end{center}

\clearpage
\begin{center}
\includegraphics[width=12cm]{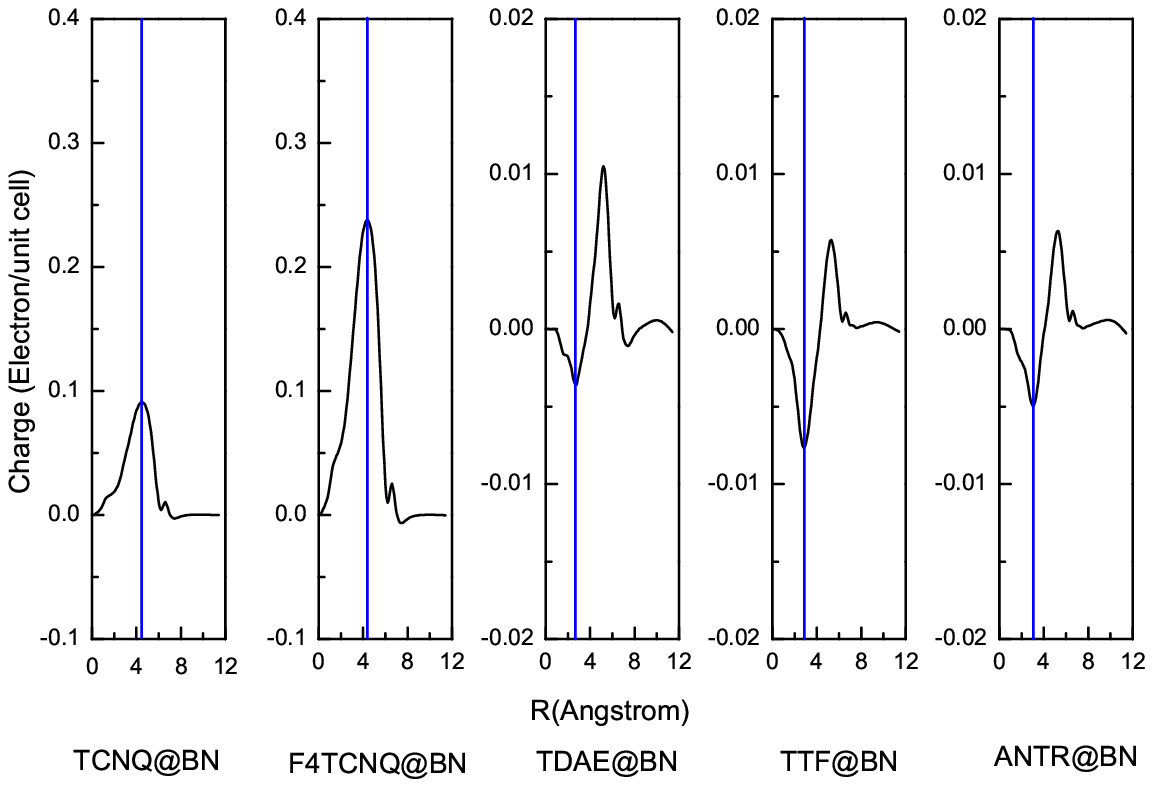} \\
\vspace{5cm} {\large He et al., JCP, Figure 4}
\end{center}

\clearpage
\begin{center}
\includegraphics[width=9cm]{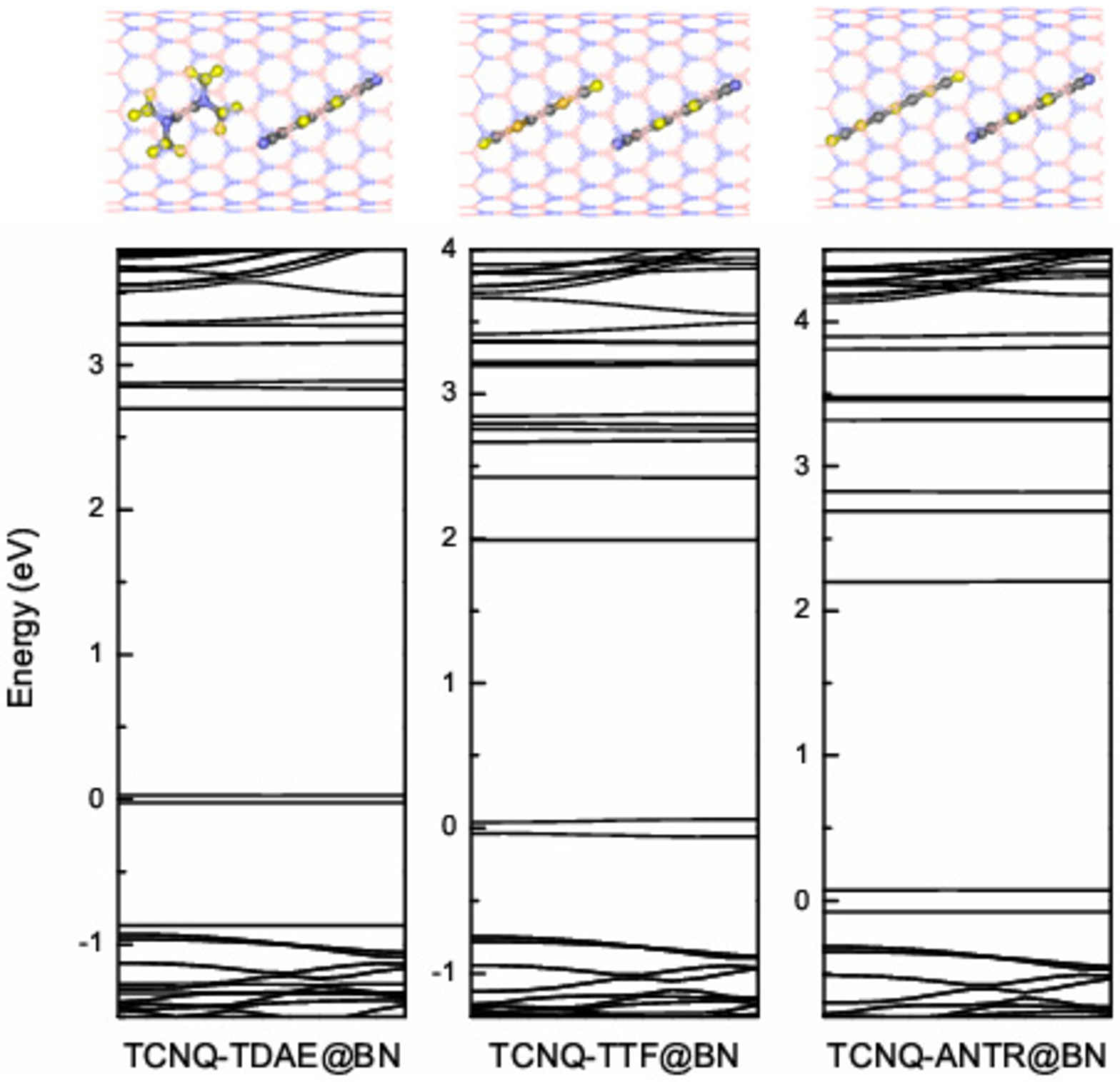} \\
\vspace{5cm} {\large He et al., JCP, Figure 5}
\end{center}

\end{document}